\documentclass[submission, Proceedings]{SciPost}

\hypersetup{
    colorlinks,
    linkcolor={red!50!black},
    citecolor={blue!50!black},
    urlcolor={blue!80!black}
}

\usepackage[bitstream-charter]{mathdesign}
\DeclareSymbolFont{usualmathcal}{OMS}{cmsy}{m}{n}
\DeclareSymbolFontAlphabet{\mathcal}{usualmathcal}

\renewcommand{\eqref}[1]{(\ref{#1})}

\newcommand{\figref}[1]{Figure~\ref{#1}}

\numberwithin{equation}{section}
\numberwithin{figure}{section}
\numberwithin{table}{section}

\begin{document}

\begin{center}{\Large \textbf{
Mulan: a part-per-million measurement of the muon lifetime and
determination of the Fermi constant \\ }}\end{center}

\begin{center}
Robert Carey\textsuperscript{1},
Tim Gorringe\textsuperscript{2$\star$} and
David Hertzog\textsuperscript{3}
\end{center}

\begin{center}
{\bf 1} Dept.\ of Physics, Boston University, Boston, MA 02215, USA
\\
{\bf 2} Dept.\ of Physics and Astronomy, University of Kentucky,
  Lexington, KY 40506, USA
\\
{\bf 3} Dept.\ of Physics, University  of Washington, Seattle, WA 98195, USA
\\
* CorrespondingAuthor@email.address
\end{center}

\begin{center}
\today
\end{center}


\definecolor{palegray}{gray}{0.95}
\begin{center}
\colorbox{palegray}{
  \begin{tabular}{rr}
  \begin{minipage}{0.05\textwidth}
    \includegraphics[width=24mm]{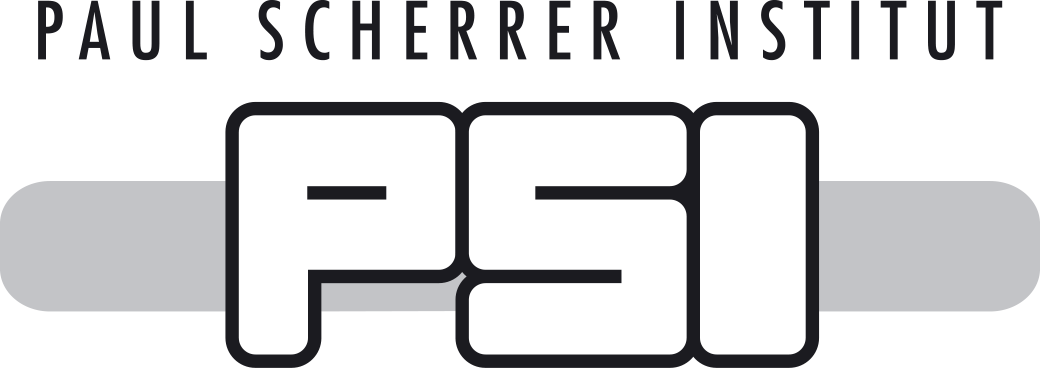}
  \end{minipage}
  &
  \begin{minipage}{0.82\textwidth}
    \begin{center}
    {\it Review of Particle Physics at PSI}\\
    \doi{10.21468/SciPostPhysProc.2}\\
    \end{center}
  \end{minipage}
\end{tabular}
}
\end{center}

\section*{Abstract}
{\bf The part-per-million measurement of the
  positive muon lifetime and determination of the Fermi constant by
  the MuLan experiment at the Paul Scherrer Institute is reviewed.  The experiment
  used an innovative, time-structured, surface muon beam and a
  near-4$\pi$, finely-segmented, plastic scintillator positron
  detector.  Two in-vacuum muon stopping targets were used: a
  ferromagnetic foil with a large internal magnetic field, and a
  quartz crystal in a moderate external magnetic field. The experiment
  acquired a dataset of $1.6 \times 10^{12}$ positive muon decays and
  obtained a muon lifetime $\tau_{\mu} = 2\, 196\, 980.3(2.2)$~ps
  (1.0~ppm) and Fermi constant G$_F = 1.166\, 378\, 7(6) \times
  10^{-5}$ GeV$^{-2}$ (0.5~ppm).  The thirty-fold improvement in
  $\tau_{\mu}$ has proven valuable for precision measurements in
  nuclear muon capture and the commensurate improvement in G$_F$ has
  proven valuable for precision tests of the standard model.}

\setcounter{section}{16}
\label{sec:mulan}

\subsection{Introduction}
\label{mulan:intro}

The electromagnetic ($\alpha_e$), strong ($\alpha_s$), gravitational
($G$) and weak ($G_F$) couplings are the ``calibration constants''
of nature \cite{Zyla:2020zbs}. Their magnitudes haven't been
determined by theory but rather are obtained from
measurement. Collectively, they determine the dynamics and bindings of
microscopic and macroscopic matter and the character of the universe.

The fine-structure constant $\alpha_e$ governs the scale of atomic
energy levels and the rates of all
electromagnetic processes.  It is known to the astonishing precision
of 0.15 parts per billion.

The energy-scale-dependent effective coupling $\alpha_s$ governs the
binding of protons and neutrons to form nuclei and the production of
chemical elements in stars.  It also controls the emergence of the two
faces of the strong interaction: quark confinement at large distances
and asymptotic freedom at short distances.  

Despite the omnipresence of the gravitational force and its
implications for the structure of the universe, the precision
determination of the gravitational constant G has been deceptively
difficult.  Since its original measurement by Cavendish, the
surprising inconsistences between modern methods have meant little
overall improvement in our knowledge of this constant
\cite{Rothleitner:2017}.

Finally, the weak interaction governs the thermonuclear reactions in
the sun that are ultimately responsible for light, energy and
life. The understanding of weak interactions enables the computation
of phenomena from cosmology and astrophysics to nuclear and particle
physics, including exacting tests of electroweak theory.  Fermi
described the weak processes by a simple four-fermion contact
interaction with the coupling strength that became known as
$G_F$. This constant and the current-current weak interaction
description have survived many decades as a very convenient,
low-energy, effective theory. Of course, our modern understanding of
weak interactions has evolved to incorporate such features as parity
violating $V-A$ currents and heavy $W, Z$ gauge bosons, in a unified
electroweak theory described by two gauge couplings and the Higgs
energy density. The Fermi constant $G_F$, together with measurements
of $\alpha_e$ and $M_Z$, provide by far the best determinations of the
gauge couplings and Higgs energy density.

Since its discovery in 1933, the muon, a heavy sibling of ordinary
electrons, has played a significant role in subatomic
physics. Muons are undoubtedly the best tool for the precise
determination of the Fermi constant and, uniquely from the
considerations above, provide by far the most precise measure of the
weak coupling.  From the theoretical perspective, the 
purely-leptonic muon decay is well suited to precision calculations within the
Fermi theory, and from the experimental perspective, its
microsecond-scale lifetime is well suited to modern techniques for
time measurements. Because the best method to determine
$G_F$ is from the muon lifetime, it is appropriate to recognize that
what is measured is $G_\mu$, the muon constant in weak decay.  The
assumption of lepton universality allows the relation $G_F \equiv
G_\mu$, which we assume here, but can and should be challenged by
other weak interaction processes.


An important breakthrough for determining G$_F$ was work by van
Ritbergen and Stuart \cite{vanRitbergen:1999fi} and Pak and Czarnecki
\cite{Pak:2008qt}. Using Fermi theory with 2-loop QED corrections,
these authors reduced the theoretical uncertainty in the relation
between the muon lifetime and the Fermi constant from
15~parts-per-million to 150~parts-per-billion.  Their work thus opened
the door for the MuLan experiment at PSI
\cite{Webber:2010zf,Tishchenko:2012ie}, a part-per-million
measurement of the muon lifetime $\tau_{\mu}$ and determination of the
Fermi constant G$_F$ -- a thirty-fold improvement over earlier
measurements.

\subsection{Experimental setup}
\label{mulan:setup}

The principle of the MuLan measurement of the muon lifetime is
straightforward.  

First, prepare a small ``source'' of positive
muons. Next, measure the times of decay positrons.  Finally, construct
the exponential decay curve and extract the positive muon lifetime.
In practice we repeated the sequence of source preparation and
positron measurement at approximately 30~KHz over a period of roughly
20 weeks in two running periods.

The experiment used longitudinally polarized,
29~MeV/c, positive muons from the $\pi$E3 secondary beamline at the PSI
proton cyclotron.
Incoming muons were stopped in solid targets and outgoing positrons
were detected in a near-4$\pi$, finely segmented, fast-timing, plastic
scintillator positron detector.  The analog signals from individual
scintillators were recorded by 450~MSPS (mega samples per second) waveform digitizers and
accumulated by a high-speed data acquisition system.

One innovative feature of the system was the imposition 
of time structure in the $\pi$E3 beamline.
The experiment operated in repeating cycles of
beam-on accumulation periods, in which surface muons
were accumulated in the stopping target, and beam-off measurement periods,
in which decay positrons were detected in the MuLan detector.
The time structure avoided the need to associate the decay positrons
with parent muons -- a limiting factor of earlier experiments
using continuous beams.

\begin{figure}[th]
\begin{center}
\includegraphics[width=0.8\textwidth]{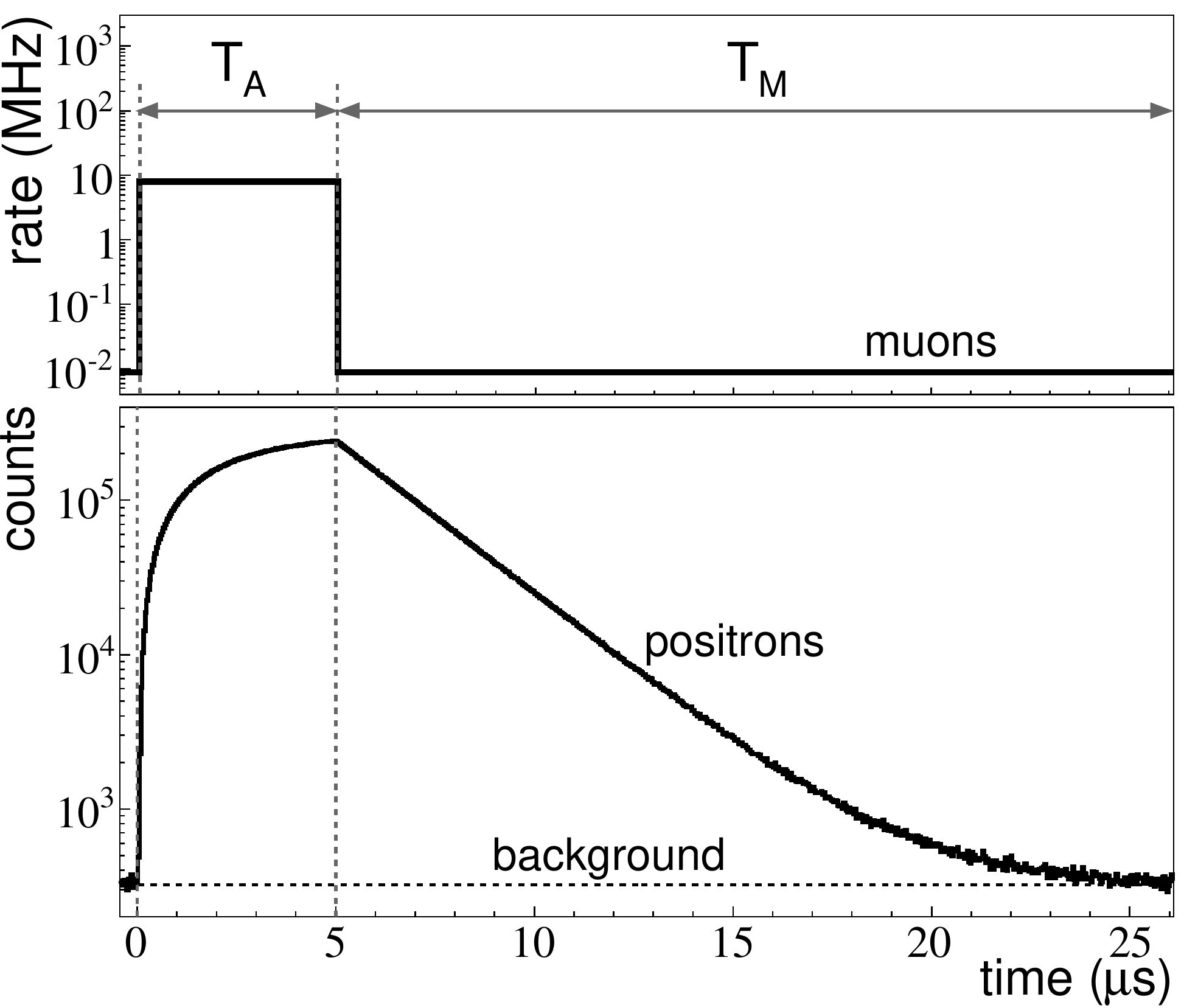}
\caption{Plot of the time dependence of the muon arrival rate (upper panel)
  and decay positron counts (lower panel) that was imposed
  by the electrostatic kicker. The durations of the beam-on accumulation period
  and beam-off measurement period were $T_A$ = 5~$\mu$s and $T_M$ = 22~$\mu$s,
  respectively. Figure courtesy of the MuLan collaboration.}
\label{fig:lifetime}
\end{center}
\end{figure}

The specific time structure comprised a 5~$\mu$s-long beam-on
accumulation period ($T_A$), and a 22~$\mu$s-long beam-off
measurement period ($T_M$).  The time structure was imposed on the $\pi$E3
beam using a custom-built, fast-switching, 25~kV electrostatic kicker.
When the kicker was de-energized, the muons were transported to the
Target; when the kicker was energized, the muons were deflected
into a collimator.  A sample time distribution of incoming muons and
outgoing positrons that illustrates the accumulation and measurement
periods is shown in \figref{fig:lifetime}.

Other innovative features of the experiment were the use of in-vacuum stopping
targets and near-4$\pi$ positron detection.  A consequence
of parity violation in weak interactions is that the emitted positrons
in muon decay are asymmetrically distributed about the muon spin
direction.  This poses a problem as spin precession and spin
relaxation of stopped muons could distort the pure exponential time
distribution of the decay positrons and bias the extraction of
$\tau_{\mu}$.

A fully 4$\pi$, perfectly isotropic, positron detector would negate
this issue of precession and relaxation by detecting positrons with
identical probability in all directions.  The Mulan combination of an
in-vacuum, detector-centered target for incoming muons and
near-4$\pi$, near-isotropic, detector for outgoing positrons, was an
important part of the experimental strategy for minimizing such spin
precession and relaxation effects.

In addition, the experiment deployed two different combinations of
stopping target materials and transverse magnetic fields in order to
further reduce the spin precession and relaxation effects.  One setup
involved a magnetized Fe-Cr-Co foil (Arnokrome-III) with a $\sim$4~kG
internal $B$-field.  Another setup involved a quartz crystal disk
(SiO$_2$) in a 130~G external $B$-field.  In the ferromagnetic target,
where muons reside as diamagnetic ions, the $\mu^+$ precession
frequency was about 50~MHz.  In the quartz target the primary muonium
($\mu^+e^-$) population has a 180~MHz precession frequency and the
secondary muon ($\mu^+$) population has a 1.8~MHz precession
frequency.  In both strategies, spin dephasing during muon
accumulation yielded a roughly 1000-fold reduction in the
ensemble-averaged $\mu^+$ polarization at the beginning of the
measurement period.

\begin{figure}
\begin{center}
\includegraphics[width=0.6\textwidth]{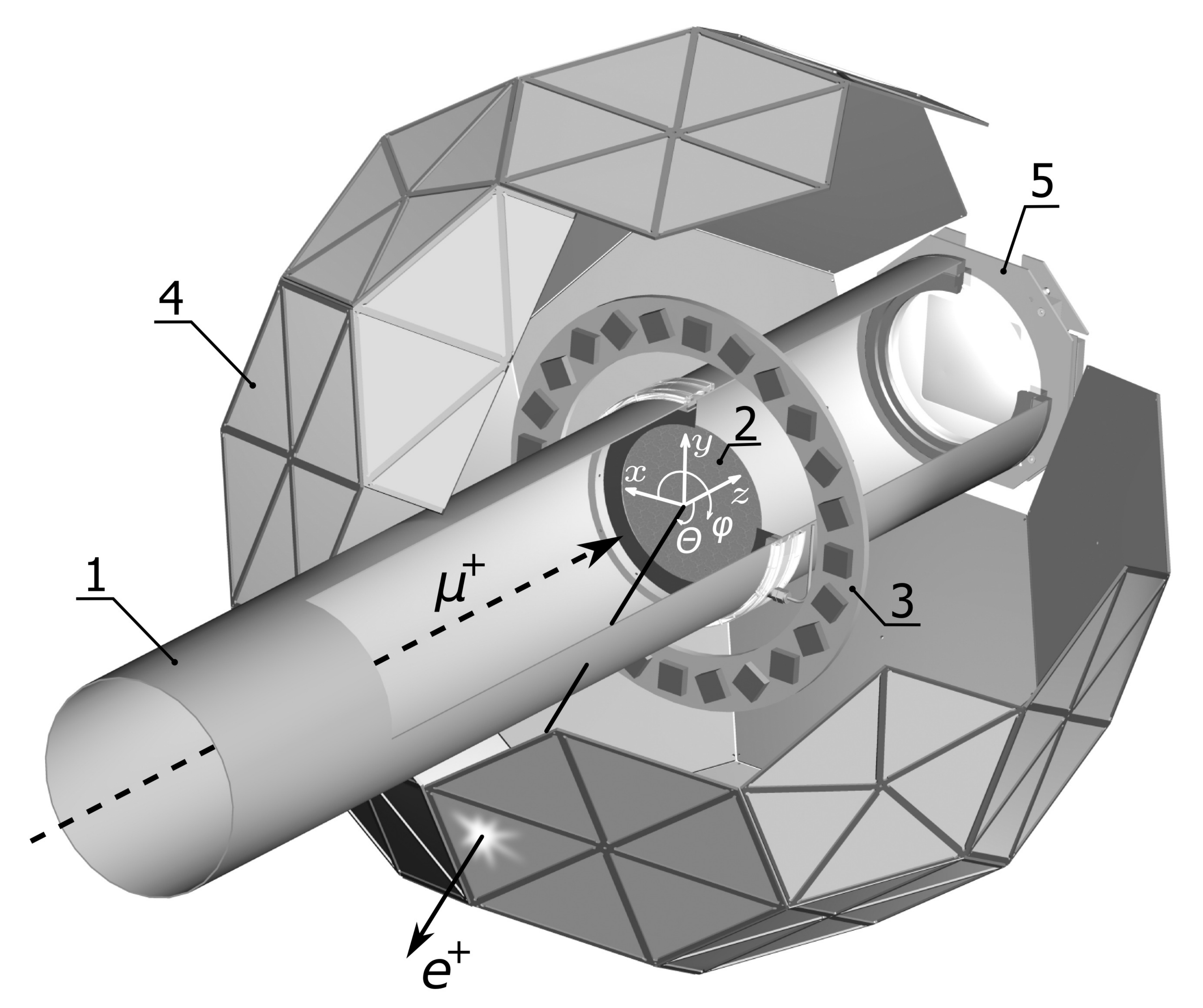}
\caption{A cutaway view of the Mulan experimental setup showing the
  (1) vacuum beamline, (2) in-vacuum stopping target, (3) Halbach
  arrangement permanent magnet, (4) soccer ball geometry scintillator
  array, and (5) beam monitor.  We used the Halbach magnet for the
  external magnetic field in the quartz target data-taking. Figure
  courtesy of the MuLan collaboration.}
\label{fig:detector}
\end{center}
\end{figure}

The positron detector was constructed of 170 triangle-shaped, plastic
scintillator pairs arranged in a soccer ball geometry
(\figref{fig:detector}).  Each pair comprised an inner and outer
scintillator tile. The pairs were grouped into ten pentagonal
enclosures containing five tile-pairs and twenty hexagonal enclosures
containing six tile-pairs, which together formed the soccer ball
geometry.  The segmentation was important in reducing positron pile-up
in individual detector elements.  The symmetric arrangement of
detector elements was important in reducing the effects of muon spin
rotation / relaxation.

\subsection{Data analysis}

A total of 1.1$\times$10$^{12}$ decays from positive muon stops in
Arnokrome-III and 5.4$\times$10$^{11}$ decays from positive muon stops
in quartz were collected.
Other datasets with different orientations of the magnetic field
and different centering of the muon stopping distribution
were collected in order to study the systematic errors associated with
spin precession and relaxation.

The time and amplitude of individual pulses were determined from least
square fits to digitized waveform templates.  The procedure fit a
higher-resolution template waveform (0.22~ns sampling-interval) to the
lower-resolution individual waveforms (2.2~ns sampling-interval).  The
higher-resolution templates were constructed by combining a large
sample of 2.2~ns sampling-interval, single positron, digitized
waveforms.  The fitting procedure would add / remove pulses to obtain
the best $\chi^2$.

Positrons were defined as inner-outer tile-pair coincidences.  In
identifying the coincidences, cuts were applied to define an
unambiguous amplitude threshold $A_{thr}$ for detector hits and
to define an unambiguous artificial deadtime
(ADT) between detector pulses.  Hits that survived these cuts were
sorted into time distributions of inner singles, outer singles and
inner-outer coincidences.  The construction of coincidence histograms
with different thresholds and deadtimes was important for studying the
distortions that arise from pulse pileup and gain changes.  The
typical rates were 40 stopped muons per accumulation period and 15
detected positrons per measurement period. The nominal 13.3~ns ADT
yielded a pileup distortion of roughly $10^{-3}$ at the start of the
measurement period and roughly $10^{-7}$ at the end of the measurement
period.

A hit is lost if it occurs in the artificial deadtime of an earlier hit. 
Our procedure for correcting for pileup took advantage of the time structure 
of the incident beam. The pileup losses were statistically recovered 
by replacing the lost hits in each measurement period with
measured hits at equivalent times in neighboring measurement periods.
For example, to correct for leading-order pileup, 
if a hit is observed at time $t_i$ in fill $j$ (the ``trigger'' hit),
a hit is searched for within the interval $t_i$~$\rightarrow$~$t_i + \mathrm{ADT}$ 
in fill $j+1$ (the ``shadow'' hit).  Adding the resulting histogram 
of shadow hit times to the original histogram of trigger hit times
thus statistically recovers the lost hits (similar shadow methods were employed for
handling the higher-order pileup).

As mentioned, only hits with amplitudes exceeding the threshold
$A_{thr}$ were used.  Consequently, if the detector gain changes over
the measurement period, then the time histogram will be distorted by
either additional hits climbing above $A_{thr}$ or additional hits
falling below $A_{thr}$ cut.  We corrected for gain changes versus
measurement time by monitoring changes in the positron minimally
ionizing particle (MIP) peak amplitude over the measurement period.

A simple procedure was used to extract the lifetime $\tau_{\mu}$ from
the Arnokrome-III target.  The summed tile-pair time histogram of
coincidence hits was fit to $N e^{-t/\tau_{\mu}} + C$.  The approach
relied on sufficient cancellation of Arnokrome-III precession and
relaxation effects by combination of the spin dephasing and the
opposite-pair detector geometry.

A more complicated procedure was needed to extract the lifetime
$\tau_{\mu}$ from the quartz target.  First, 170 geometry-dependent
effective lifetimes were extracted for each tile-pair from fits to
\begin{equation}
N(t) = N e^{-t/\tau_\mathrm{eff}} [ 1 + f(t) ] + C ,
\label{eq:Lambda:d}
\end{equation}
where $f(t)$ accounts for time-dependent effects of transverse-field
(TF) spin precession and relaxation.  Then, the true positive muon
lifetime $\tau_{\mu}$ was extracted by fitting the effective
lifetimes, $\tau_\mathrm{eff}$, to
\begin{equation}
\tau_\mathrm{eff} ( \theta_B , \phi_B  ) = \tau_{\mu} [ 1 +  \delta ( \theta_B , \phi_B ) ]  ,
\label{eq:fitfunction:sophisticated}
\end{equation}
where $\delta ( \theta_B , \phi_B )$ accounts for geometry-dependent
effects of longitudinal-field (LF) spin relaxation.  Together the two
steps were sufficient to handle the effects of precession and
relaxation in quartz.


\subsection{Results}

The individual results from the Arnokrome-III dataset and the quartz dataset, and the weighted average are given in Table \ref{mulan:results}.

\begin{table}[ht]
\begin{center}
\begin{tabular}{|l|c|}

\hline 
 Target material & Positive muon lifetime (ps) \\
\hline
Arnokrome-III  &  $2~196~979.9 \pm 2.5 (stat) \pm 0.9 (syst)$ \\
Quartz & $2~196~981.2 \pm 3.7 (stat) \pm 0.9 (syst)$ \\
\hline
Weighted average & $2~196~980.3 \pm  2.1 (stat) \pm 0.7 (syst)$ \\
\hline
\end{tabular}
\end{center}
\caption{Muon lifetime results from the Arnokrome-III dataset, quartz dataset,
and their weighted average. \label{mulan:results}}
\end{table}
  
The weighted average corresponds to an overall uncertainty in the
positive muon lifetime of 2.2~ps, or 1.0~ppm. The largest
contributions to the systematic uncertainties are associated with the
aforementioned pulse pileup, gain changes, and muon precession and
relaxation effects, as well as the knowledge of the time independence
of the beam extinction during the measurement period.  The final
result for $\tau_{\mu}$ is in agreement with the earlier work of
Giovanetti {\it et al.} \cite{Giovanetti:1984yw}, Balandin {\it et al.}
\cite{Balandin:1975fe} and Bardin {\it et al.} \cite{Bardin:1984ie}.

We note the precision determination of $\tau_{\mu}$ is important to
work on nuclear muon capture. The MuCap experiment \cite{Andreev:2007wg} at PSI
determined the $\mu$p singlet capture rate from the small difference
between the positive muon lifetime and the muonic hydrogen atom
lifetime. Similarly, the MuSun experiment \cite{Kammel:2013ara} at PSI will
determine the $\mu$d doublet capture rate from the small difference
between the positive muon lifetime and the muonic deuterium atom
lifetime. These two experiments are described in Section~17~\cite{section17} and Section~18~\cite{section18}, respectively. 

The Fermi constant G$_F$ was extracted using the relation obtained by
van Ritbergen and Stuart (vRS) \cite{vanRitbergen:1999fi} and yields
$G_F ({\rm MuLan}) = 1.166\, 378\, 7(6)\times 10^{-5}$GeV$^{-2}
(0.5~\mathrm{ppm})$ -- a thirty-fold improvement over the earlier 1998
Particle Data Group \cite{Caso:1998tx} value that pre-dated the vRS
theoretical work and MuLan experimental work.  The 0.5~ppm error is
dominated by the 1.0~ppm uncertainty of the lifetime measurement, with
contributions of 0.08~ppm from the muon mass measurement and 0.15~ppm
from the theoretical corrections.

Together, the fine structure constant $\alpha$, Fermi coupling
constant G$_F$, and Z boson mass M$_Z$, fix the electroweak parameters
of the standard model. The thirty-fold improvement in the
determination of the Fermi constant G$_F$, together with other
improvements in determinations of $\alpha$ and M$_Z$, have allowed for
improved tests of the standard model and improved searches for new
phenomena.

We wish to thank our collaborators in the MuLan experiment
and Paul Scherrer Institute for their exceptional organizational
and technical support. We also wish to thank  M.~ Barnes and G.~Wait
from TRIUMF for their development of the electrostatic kicker,
Bill Marciano for advocating and promoting the experiment, 
and the U.S.\ National Science Foundation (NSF 1807266)
and U.S.\ Department of Energy (DOE DE-FG02-97ER41020) 
for their financial support.

\bibliography{mulan}

\begin{thebibliography}{10}
\providecommand{\url}[1]{\texttt{#1}}
\providecommand{\urlprefix}{URL }
\expandafter\ifx\csname urlstyle\endcsname\relax
  \providecommand{\doi}[1]{doi:\discretionary{}{}{}#1}\else
  \providecommand{\doi}{doi:\discretionary{}{}{}\begingroup
  \urlstyle{rm}\Url}\fi
\providecommand{\eprint}[2][]{\url{#2}}

\bibitem{Zyla:2020zbs}
P.~Zyla \emph{et~al.},
\newblock \emph{{Review of Particle Physics}},
\newblock PTEP \textbf{2020}(8), 083C01 (2020),
\newblock \doi{10.1093/ptep/ptaa104}.

\bibitem{Rothleitner:2017}
C.~Rothleitner and S.~Schlamminger,
\newblock \emph{Measurements of the newtonian constant of gravitation},
\newblock Rev. Sci. Instrum. \textbf{88}, 111101 (2017),
\newblock \doi{10.1063/1.4994619}.

\bibitem{vanRitbergen:1999fi}
T.~van Ritbergen and R.~G. Stuart,
\newblock \emph{{On the precise determination of the Fermi coupling constant
  from the muon lifetime}},
\newblock Nucl.Phys. \textbf{B564}, 343 (2000),
\newblock \doi{10.1016/S0550-3213(99)00572-6},
\newblock \eprint{hep-ph/9904240}.

\bibitem{Pak:2008qt}
A.~Pak and A.~Czarnecki,
\newblock \emph{{Mass effects in muon and semileptonic $b\to c$ decays}},
\newblock Phys. Rev. Lett. \textbf{100}, 241807 (2008),
\newblock \doi{10.1103/PhysRevLett.100.241807},
\newblock \eprint{0803.0960}.

\bibitem{Webber:2010zf}
D.~Webber \emph{et~al.},
\newblock \emph{{Measurement of the Positive Muon Lifetime and Determination of
  the Fermi Constant to Part-per-Million Precision}},
\newblock Phys. Rev. Lett. \textbf{106}, 041803 (2011),
\newblock \doi{10.1103/PhysRevLett.106.041803},
\newblock \eprint{1010.0991}.

\bibitem{Tishchenko:2012ie}
V.~Tishchenko \emph{et~al.},
\newblock \emph{{Detailed Report of the MuLan Measurement of the Positive Muon
  Lifetime and Determination of the Fermi Constant}},
\newblock Phys. Rev. D \textbf{87}(5), 052003 (2013),
\newblock \doi{10.1103/PhysRevD.87.052003},
\newblock \eprint{1211.0960}.

\bibitem{Giovanetti:1984yw}
K.~Giovanetti \emph{et~al.},
\newblock \emph{{Mean Life of the Positive Muon}},
\newblock Phys. Rev. D \textbf{29}, 343 (1984),
\newblock \doi{10.1103/PhysRevD.29.343}.

\bibitem{Balandin:1975fe}
M.~Balandin, V.~Grebenyuk, V.~Zinov, A.~Konin and A.~Ponomarev,
\newblock \emph{{Measurement of the Lifetime of the Positive Muon}},
\newblock Sov. Phys. JETP \textbf{40}, 811 (1975),
\newblock Available at
  \url{http://ivanik3.narod.ru/TimeLifeMezon/e_040_05_0811Balandin.pdf}.

\bibitem{Bardin:1984ie}
G.~Bardin, J.~Duclos, A.~Magnon, J.~Martino, E.~Zavattini, A.~Bertin,
  M.~Capponi, M.~Piccinini and A.~Vitale,
\newblock \emph{{A New Measurement of the Positive Muon Lifetime}},
\newblock Phys. Lett. B \textbf{137}, 135 (1984),
\newblock \doi{10.1016/0370-2693(84)91121-3}.

\bibitem{Andreev:2007wg}
V.~Andreev \emph{et~al.},
\newblock \emph{{Measurement of the rate of muon capture in hydrogen gas and
  determination of the proton's pseudoscalar coupling g(P)}},
\newblock Phys. Rev. Lett. \textbf{99}, 032002 (2007),
\newblock \doi{10.1103/PhysRevLett.99.032002},
\newblock \eprint{0704.2072}.

\bibitem{Kammel:2013ara}
P.~Kammel,
\newblock \emph{{Precision Muon Capture at PSI}},
\newblock PoS \textbf{CD12}, 016 (2013),
\newblock \doi{10.22323/1.172.0016}.

\bibitem{section17}
M.~Hildebrandt and C.~Petitjean,
\newblock \emph{{MuCap: Muon Capture on the Proton}},
\newblock SciPost Phys. Proc. \textbf{2}, ppp (2021),
\newblock \doi{10.21468/SciPostPhysProc.2.XXX}.

\bibitem{section18}
P.~Kammel,
\newblock \emph{{MuSun - Muon Capture on the Deuteron}},
\newblock SciPost Phys. Proc. \textbf{2}, ppp (2021),
\newblock \doi{10.21468/SciPostPhysProc.2.XXX}.

\bibitem{Caso:1998tx}
C.~Caso \emph{et~al.},
\newblock \emph{{Review of particle physics. Particle Data Group}},
\newblock Eur. Phys. J. C \textbf{3}, 1 (1998),
\newblock \doi{10.1007/s10052-998-0104-x}.

\end{thebibliography}

\nolinenumbers

\end{document}